\documentstyle[aps,prl,twocolumn,psfig]{revtex}
\begin{document}
\draft
\title{Intrinsic vs. laboratory frame description of the deformed
  nucleus $^{\bf 48}$Cr.}
\author{E. Caurier$^{1)}$, J.L. Egido$^{2)}$,
  G. Mart\'{\i}nez-Pinedo$^{2)}$, A. Poves $^{2)}$, J. Retamosa$^{2)}$,
  L.M. Robledo$^{2)}$ and A.P. Zuker$^{1,2)}$}

\address{$^{1)}$
 Physique Th\'eorique. B\^at 40/1
Centre de Recherches Nucleaires
BP28 F-67037 Strasbourg Cedex-2 , France}
\address{$^{2)}$
Departamento de F\'\i sica Te\'orica C-XI.
Universidad Aut\'onoma de Madrid
E--28049 Madrid, Spain}
\maketitle

\begin{abstract}
The collective yrast band of the nucleus $^{48}$Cr is studied using
the spherical shell model and the HFB method. Both approaches produce
basically the same axially symmetric intrinsic state up to the
- accurately reproduced - observed backbending.
Agreement between both calculations extends to most observables. The
only significant discrepancy comes from the static moments of inertia
and can be attributed to the need of a more refined treatment of
pairing correlations in the HFB calculation.
\end{abstract}


The study of the collective behavior of deformed nuclei is a classical
problem in Nuclear Physics. Traditionally, mean field descriptions in
the intrinsic frame have been favoured, as they take naturally
advantage of the spontaneous breakdown of rotational symmetry. The
price to pay for the gain in physical insight is the loss of angular
momentum as good quantum number.

In the laboratory frame description, as provided by spherical
shell model calculations (SM), angular momentum is conserved but
the physical insight, associated to the existence of an intrinsic
state is lost, except in the very rare cases where Elliott's SU3
symmetry \cite{Elliott} operates. Furthermore, the approach
suffers form numerical limitations. Hence, so far, it had been
implemented mostly  in regions such as  the $p$ and $sd$
shells where the number of active particles is too small for
collective features to become dominant. Nonetheless, there are
a few nuclei - such  as $^{20}$Ne and $^{24}$Mg - that are well
 reproduced by the SM calculations and do exhibit collective
properties, whose origin can be traced to the approximate validity
 of the SU3 symmetry, for which the relationship between
the intrinsic and laboratory frame descriptions is well understood.

In regions where the SU3 symmetry is poorly respected, as in
in the $pf$ shell \cite{BMcG}, the study of potentially good
``rotors'' was impaired by lack of experimental evidence, and by
the difficulty of an exact SM treatment beyond 5 active  particles.
The situation has changed through recent measurements \cite{CAM94}
demonstrating  that  $^{48}$Cr is a good rotor up to spin
 $J=10 $ where the yrast band bends back. This behavior is
reminiscent of the situation in much heavier deformed nuclei.
Simultaneously, full $pf$ calculations  \cite{pfcal} have
become available, that reproduce in detail the observed properties
of A=48 isobars, and in particular those of $^{48}$Cr.

Therefore, this nucleus provides an unique testing ground to compare
the SM (laboratory frame) description of permanent deformation with
Cranked Hartree- Fock- Bogoliubov (CHFB) calculations \cite{CHFBcal}
with the finite range density dependent Gogny force \cite{Gogny};
which represent the (self-consistent) state of the art formulation of
the intrinsic frame approach.

{}From the comparison it should be possible to obtain a better
understanding of the intrinsic structure of the SM solutions,
which in turn, may indicate in what sense the CHFB description
falls short of an exact one.

\bigskip

{\bf Computational procedures}.
In the Spherical Shell Model (SM) $^{48}$Cr is described in a $0\hbar
\omega$ space, i.e. eight particles are allowed to occupy all the
 states available in the $pf$ shell (1963461 states). The effective
interaction is
given by a minimally modified version of the Kuo-Brown's G-matrix
\cite
{MSSMint} denoted KB3 in \cite{pfcal}. The single particle energies
are
taken from the $^{41}$Ca experimental spectrum. The effect of core
polarization on the quadrupole properties is taken into account by the
use
of effective charges $q_\pi =1.5$, $q_\nu =0.5$. The Hamiltonian is
treated
by the Lanczos method and diagonalized by the code ANTOINE
\cite{ANTOINE}.

In the intrinsic frame calculations we have used the Self Consistent
Cranking Hartree- Fock- Bogoliubov method ( CHFB) with the density
dependent
Gogny force. The CHFB equations determining the mean field intrinsic
state $
|\phi _\omega \rangle $ are obtained by imposing the condition that
the mean value of the Routhian be stationary against small variations
of the intrinsic state, i.e.,
\begin{equation}
\delta \langle \phi _\omega |\hat H-\omega \hat J_x-\lambda _N\hat
N-\lambda
_Z\hat Z|\phi _\omega \rangle =0.  \label{SCCeq}
\end{equation}
The Lagrange multipliers $\omega $, $\lambda _N$ and $\lambda _Z$ are
determined by the usual angular momentum and particle number
constraints
$\langle \phi _\omega |\hat J_x|\phi _\omega \rangle =\sqrt{I(I+1)}$,
$\langle \phi _\omega |\hat N|\phi _\omega \rangle =N$ and $\langle
\phi_\omega |\hat Z|\phi _\omega \rangle =Z$.

The HFB wave functions have been expanded in a triaxial harmonic
oscillator basis $|n_xn_yn_z\rangle $ with different oscillator
lengths. Ten oscillator shells are included in order to ensure the
convergence of the mean field results.  The parameters of the Gogny
force used in this calculation were adjusted more than ten years ago
to reproduce ground state bulk properties of nuclei (DS1 set
\cite{GognyPar}).  Without further changes, this force has proven
capable of describing successfully many phenomena, and in particular
high spin behaviour \cite{CHFBcal}.

In order to understand more qualitatively the physics involved and to
make contact with the Shell Model calculations
we have computed the following quantities in a spherical
representation of the basis:

The ``fractional shell occupancy''
\begin{equation}
\nu (n,l,j)=\frac 1{2j+1}\sum_{m=-j}^{m=j}\langle \phi _\omega
|c_{nljm}^{+}c_{nljm}|\phi _\omega \rangle ,
\label{occshell}
\end{equation}
the ``shell contribution to $\langle J_x\rangle $''

\begin{equation}
  \begin{array}{l}
    \displaystyle{j_x(n,l,j)=}\\
    \displaystyle{\sum_{m,m^{\prime }}(J_x)_{(nljm),(nljm^{\prime })}\langle
    \phi_\omega |c_{nljm}^{+}c_{nljm^{\prime }}|\phi _\omega \rangle }
  \end{array}
  \label{jxshell}
\end{equation}
and the ``shell contribution to the quadrupole moment''

\begin{equation}
  \begin{array}{l}
\displaystyle{Q_{20}(n,l,j;n^{\prime },l^{\prime },j^{\prime })=} \\
\displaystyle{\sum_{m,m^{\prime
}}(q_{20})_{(nljm),(n^{\prime }l^{\prime }j^{\prime }m^{\prime
})} \langle
\phi _\omega |c_{nljm}^{+}c_{n^{\prime }l^{\prime }j^{\prime
    }m^{\prime
}}|\phi _\omega \rangle.}
  \end{array}
\label{q20shell}
\end{equation}
In the above formulae $|\phi _\omega \rangle $ is the intrinsic CHFB
wave
function expressed in the triaxial basis and $c_{nljm}^{+}$ are the
operators creating a particle in the harmonic oscillator orbit
$|nljm\rangle$
with oscillator length $b_0=(b_xb_yb_z)^{1/3}$. In order to obtain
these
quantities the triaxial basis has been expanded in a spherical one
following a procedure similar to that of Ref. \cite{chaswal}. As the
triaxial basis has, in general, different oscillator lengths the
expansion contains in principle an  infinite number of terms.
 In our case, an efficient truncation is obtained by allowing
the spherical basis  to contain four major shells beyond those in the
triaxial basis. The convergence of the truncation has been checked by
comparing $\sum_{nlj}(2j+1)\nu (nlj)$, $\sum_{nlj}j_x(nlj)$ and $
\sum_{nlj;n^{\prime }l^{\prime }j^{\prime }}q_{20}(nlj;n^{\prime
  }l^{\prime
}j^{\prime })$  with $\langle N\rangle $, $\langle
J_x\rangle $ and $\langle Q_{20}\rangle $ respectively. The
differences are
typically of the order of $0.01\,\%$.

\bigskip

{\bf Energetics}.In Fig. 1 the SM,  CHFB  and experimental gamma ray
energies $
E_\gamma (J)=E(J)-E(J-2)$ are plotted as a function of the angular
momentum
$J$. The SM results nicely reproduce the experiment including the
backbending
seen at $J=10$. On the other hand, the mean field values of
$E_\gamma $
follow the same trend as the experimental and SM ones but they are
shifted
downwards by $\approx 0.5$ MeV. This means that the mean field dynamic
moment
of inertia (${\cal J}^{(2)}(J)=4/\Delta E_\gamma $) is similar to the
SM and
experimental ones although the static moment of inertia (${\cal J}
^{(1)}(J)=(2J-1)/E_\gamma $ ) is on the average a factor 1.5
bigger. (The origin of this discrepancy will be explained later.)

{\bf Quadrupole properties}. The striking similarity between the SM
and CHFB results up to the backbend can be gathered from the lower
part of fig.~ 2, in which the intrinsic quadrupole moment is plotted
along the yrast band. The SM values are extracted from the BE2 values,
assuming $K=0$. The existence of an intrinsic state common to the
members of the band can be guessed directly by calculating the
contribution of a given configuration to each SM wave function,
 ({\it i.e.}, by summing the square of the amplitudes of
all basic states having the same number of particles in each
subshell). {\em These contributions are practically identical in all
  the eigenstates up to $J=10$}.  At higher spins rapid changes occur,
and the configuration in which all the particles are in the $f_{7/2}$
orbit becomes increasingly dominant. It is clear that the intrinsic
state is becoming $J$-dependent at the backbending region, and the
discrepancies in fig.~2 beyond $J=10$, suggest that it is no longer
possible to extract an intrinsic $Q_0$ from the SM results assuming a
$K=0$ band. In the upper part of the figure an alternative is
proposed, by comparing the B(E2) values, obtained directly in the SM
case with those derived from CHFB by applying the generalization of
the rotational model prescription to small triaxialities (see
\cite{CHFBcal}). The agreement is again nearly perfect up to $J=10$
but then deteriorates, although not as much as in the lower figure.

In assessing the significance of these results  we should
keep in mind that they are in both cases (rotational) model dependent.
They indicate that the model is as good as exact up to the backbend,
and then breaks down - at least in the standard implementation
proposed here. They certainly {\em do not}
indicate that the SM and CHFB descriptions are becoming different.
On the contrary, we shall find  evidence of their closeness.

{\bf Orbital occupancies}. In figure 3 are have plotted the fractional
occupancies of the spherical orbits in the CHFB solution (eq.
(\ref{occshell}
)) (upper part) and in the SM one (lower part). In all cases they
 are quite constant up to the backbending where the
$f_{7/2}$ orbit becomes rapidly the only relevant one.

{\em However\/}: the $f_{7/2}$ occupancy is {\em always} the largest
by far , and in the CHFB case the contribution $j_x(f_{7/2})$ to
$\langle J_x\rangle$ in eq.(\ref{jxshell}) is always greater than
 99 \%.
It means that the $f_{7/2}$ orbit plays a major part in the
two yrast regimes: below backbend as the major contributor to the
deformed wavefunctions; and above through the $f_{7/2}^{8}$
configuration
that becomes increasingly dominant. This picture is consistent with
the usual idea that the backbend is associated with alignment of
$f_{7/2}$ particles, which are also massively present in the
collective regime at low spin.

{\bf Magnetic properties}. In Figure 4 we present the
CHFB and SM results for the gyromagnetic factor $g$. In both
cases and up to the backbending zone they are close to the rotational
limit $g_R=Z/A=0.50$. For a pure $f_{7/2}$ configuration the value of
 $g$ is also
constant and equal to 0.55 explaining the slight increase in $g$ as we
enter the backbending region where these configurations become
dominant.

{\bf Pairing properties}.
{}From all we have said, it follows that the SM and CHFB results are
basically the same, except for a difference in the static moment of
 inertia . Its origin can be understood by redoing the SM calculations
 reducing the $JT=01$ two-body matrix elements involving orbits $r$
 and $t$ according to
 \begin{equation}
   \label{p}
W^{01}_{rrtt} \longrightarrow
W^{01}_{rrtt} + 0.165\sqrt{(j_r+1/2)(j_t+1/2)},
\end{equation}
which amounts to subtracting a standard pairing term ($j_r$ is the
angular momentum of orbit $r$).  The resulting $E_\gamma$ pattern for
an exact calculation with the modified interaction is shown as SM(E)
in fig.~5. To gain further insight we have also calculated in SM(P)
the energies by taking expectation values of the modified interaction
(\ref{p}) using the SM wavefunctions obtained with KB3.  (The
coefficient 0.165 was chosen - somewhat arbitrarily - to make the
first point coincide for CHFB and SM(P)). The conclusion is that:

{\em Although the energetics of the yrast band are strongly affected
by the pairing modifications, the other properties are not, since the
wavefunctions change little}.  (The overlaps $<SM(E),J|\;SM,J>$
exceed 0.97 in all cases).

The large static moments of inertia obtained in the CHFB calculations
should be attributed to an inadequate treatment of pairing effects
in a weak correlation regime: Exploratory tests using the
Lipkin-Nogami approach on top of the CHFB scheme suggest that it is
not the Gogny force that is responsible for the discrepancies but the
limitations of the mean field treatment.

{\bf $^{48}$Cr as axial rotor }
It has been recently argued \cite{rotation} that the building blocks
of wavefunctions describing good rotors are constructed by allowing
particles to move in spaces defined by  $\Delta j=2$
sequences of major shell orbits, starting on the one with  the
 largest $j$.
For these blocks, an approximate form of SU3 symmetry is valid
(quasi-SU3). One of the predictions of this model is that $^{48}$Cr
is an axially symmetric rotor, contrary to what happens to its
counterpart in the $sd$ shell, $^{24}$Mg, that obeys Elliott's SU3
and is triaxial. Experimentally no second $2^{+}$ state is  found in
$^{48}$Cr at low  excitation energy, while in $^{24}$Mg  the second
$2^{+}$ is degenerate with the yrast $4^{+}$.
In figure 6 we present the values of the deformation
parameters $\beta $ and $\gamma $ coming from the CHFB calculation.

At first, $\beta $
stays constant at $\beta \approx 0.3$, while $\gamma \approx 0$
which means that $^{48}Cr$ behaves indeed as an axial rotor
 up to the backbend. Above it, as $\beta$ decreases fast and the
system moves to a spherical regime making it difficult to interpret
in a simple way the $\gamma$ behaviour.

{\bf Effective charges}. Finally, we can separate from the total
quadrupole moment $Q_{20}$ in CHFB, the valence contribution
$Q_{20}pf$(HO) by summing \mbox{ $q_{20}(n,l,j;n^{\prime },l^{\prime
    },j^{\prime })$} in eq.~(\ref{q20shell}) over the $0f$ and $1p$
orbits, {\it i.e.}, by identifying the valence orbits with harmonic
oscillator ones.  The ratio
\[Q_{20}/Q_{20}pf({\rm HO})=1.99(J=0)\cdots 1.83(J=14) \]
is quite consistent with the isoscalar effective charge used in the SM
calculations \mbox{$q_{\nu}+q_{\pi}=2$}.

Alternatively, we can define $Q_{20}pf$(HF) by summing over all the
values of $l$, $j$ and $l^{\prime }$, $j^{\prime }$ corresponding to
the $pf$ shell, which amounts to use spherical HF orbits. This choice
naturally reduces the effective charges but they remain quite
constant since
 \[Q_{20}/Q_{20}pf({\rm HF})=1.70(J=0)\cdots 1.63(J=14). \]

\bigskip
{\bf Conclusions.}
The quantitative equivalence of the SM and CHFB descriptions has two
direct and welcome consequences:
\begin{itemize}
\item It suggests that the Gogny force must be reasonably close to
the realistic ones, consistent with NN data and known to
yield high quality spectroscopy once their bad monopole
properties are corrected.
\item It confirms the validity of the SM choice of a model space
 restricted to orbits in the vicinity of the Fermi level.
\end{itemize}
Clearly there is much to be gained by combining the simplicity and
rigour of CHFB with the SM precision and generality.

\bigskip

This work has been partially supported by the DGICyT, Spain under
grants
PB93-263 and PB91--0006, and by the IN2P3-(France) CICYT (Spain)
agreement.

% --------------------------------------------------------- REFERENCES

\begin{figure}
  \begin{center}
    \leavevmode
   \psfig{file=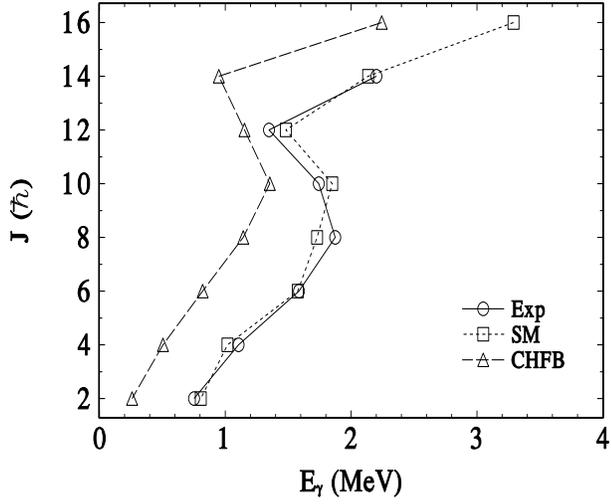,height=6.5cm,width=8.5cm}

\medskip

\caption{Yrast energies $E_{\gamma}=E(J)-E(J-2)$.}
  \end{center}
\end{figure}

\begin{figure}
  \begin{center}
    \leavevmode
    \psfig{file=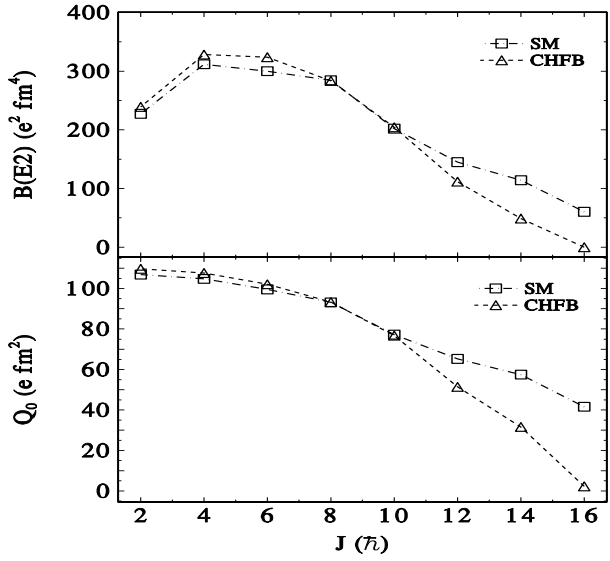,height=6.5cm,width=8.5cm}
  \end{center}
\caption{Comparing $B(E2)$ and $Q_0$ trends.}
\end{figure}

\begin{figure}
  \begin{center}
    \leavevmode
 \psfig{file=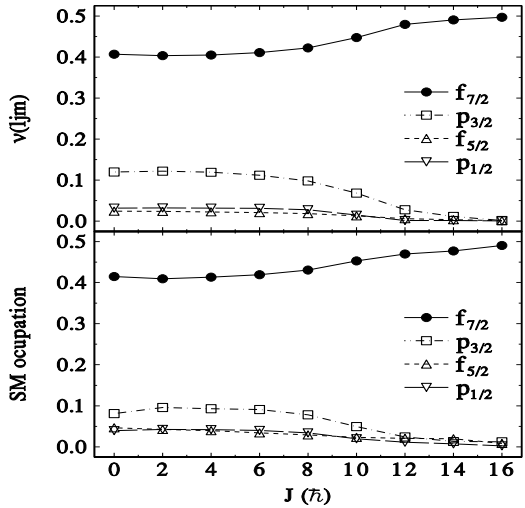,height=6.5cm,width=8.5cm}
  \end{center}
\caption{Orbital occupancies.}
\end{figure}

\begin{figure}
  \begin{center}
    \leavevmode
     \psfig{file=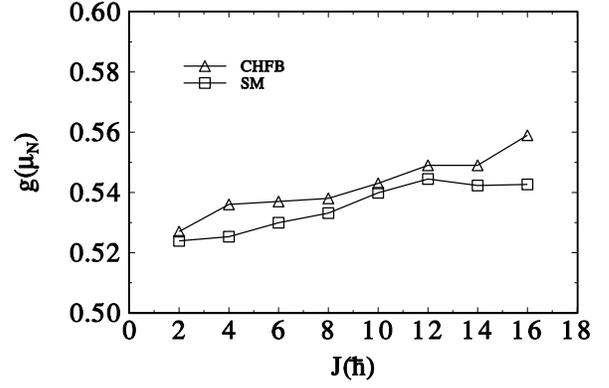,height=6.5cm,width=8.5cm}
  \end{center}
\caption{Gyromagnetic ratios}
\end{figure}

\begin{figure}
  \begin{center}
    \leavevmode
 \psfig{file=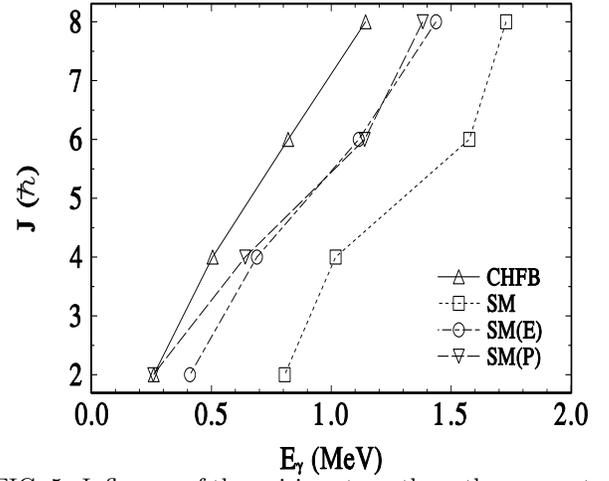,height=6.5cm,width=8.5cm}
  \end{center}
\caption{Influence of the pairing strength on the moment of
  inertia. See text.}
\end{figure}

\begin{figure}
  \begin{center}
    \leavevmode
 \psfig{file=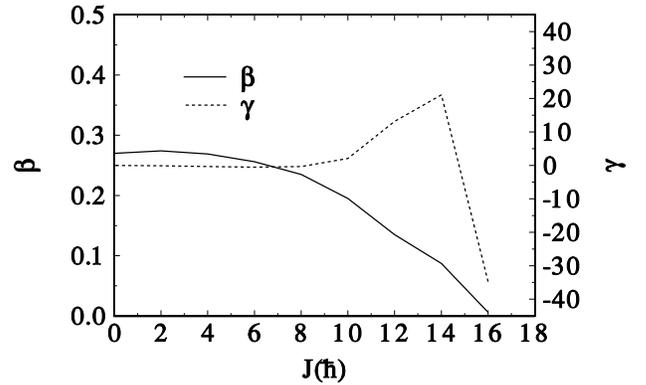,height=6.5cm,width=8.5cm}
  \end{center}
\caption{The CHFB deformation parameters.}
\end{figure}
\end{document}